\documentclass[12pt]{article}
\usepackage{epsfig,amssymb}
\renewcommand{\theequation}{\thesection.\arabic{equation}}

\begin{document}

\begin{flushright}
{KIAS-P03045} \\
{MIT-CTP-3387} \\
{\tt hep-th/0307184}
\end{flushright}

\vspace{5mm}

\begin{center}
{{{\Large \bf Tachyon Tube and Supertube
}}\\[14mm]
{Chanju Kim${}^{1}$,~~Yoonbai Kim${}^{2,3}$,~~O-Kab Kwon${}^{2}$, ~and
~Piljin Yi${}^{4}$}\\[6mm]
{\it ${}^{1}$Department of Physics, Ewha Womans University,
Seoul 120-750, Korea}\\
{\tt cjkim@ewha.ac.kr}\\[3mm]
{\it ${}^{2}$BK21 Physics Research Division and Institute of
Basic Science,\\
Sungkyunkwan University, Suwon 440-746, Korea}\\
{\tt yoonbai@skku.ac.kr~~okwon@newton.skku.ac.kr}\\[3mm]
{\it ${}^{3}$Center for Theoretical Physics,
Massachusetts Institute of Technology}\\
{\it Cambridge MA 02139, USA}\\[3mm]
{\it ${}^{4}$School of Physics, Korea Institute for Advanced Study,\\
207-43, Cheongryangri-Dong, Dongdaemun-Gu, Seoul 130-012, Korea}\\
{\tt piljin@kias.re.kr}
}
\end{center}
\vspace{10mm}

\begin{abstract}
\noindent
We search for tubular solutions in unstable D3-brane.
With critical electric field $E=1$, solutions representing supertubes,
which are supersymmetric bound states of fundamental strings, D0-branes,
and a cylindrical D2-brane, are found and shown to exhibit BPS-like
property. We also point out that boosting such a {\it tachyon tube} solution
generates string flux winding around the tube, resulting in helical
electric fluxes on the D2-brane. We also discuss issues related to fundamental
string, absence of magnetic monopole, and finally more tachyon tubes
with noncritical electric field.

\end{abstract}

\newpage

\setcounter{equation}{0}
\section{Introduction}

Supertube is a tubular D2-brane that preserves 1/4 of
supersymmetry~\cite{Ma01qs,tube,sgtube}. This requires presence of D0
and fundamental string charge, both realized as electromagnetic field on
the D2
worldvolume. Alternatively, it may be considered as  blow-up of F1-D0
composites obtained via angular momentum. One unusual
characteristic of supertube is that, as classical solution, they come
with huge degeneracy and can take many different form \cite{tube,Mateos,Ch01ys}.
For instance, the cross section of a supertube need not be circular and in fact
can even be noncompact \cite{karch}.

The simplest of such supertubes involves
a cylinder of D2 with uniform magnetic field $B$
and uniform electric field $E$ directed along the length of the
cylinder \cite{Ma01qs}.
Let us consider the field strength of type,
\begin{equation}
F= BR \,d \theta\wedge dz +E \,dt\wedge dz ,
\end{equation}
where $R$ is the radius of the cylinder while $\theta$ is of period
$2\pi$. With Born-Infeld action, the physical electric flux, along
$z$ direction and per unit $\theta$, is
\begin{equation}
\Pi_z= R\frac{E}{\sqrt{1-E^2+B^2}},
\end{equation}
and the energy density per unit $z$ and unit $\theta$ is
\begin{equation}
{\cal E}=R\frac{1+B^2}{\sqrt{1-E^2+B^2}} =\sqrt{(\Pi_z^2+R^2)(1+B^2)}\, .
\end{equation}
Note that the total electric flux is $2\pi \Pi_z$ while the net
magnetic flux per unit $z$ is $2\pi R B$.

This configuration is known to become 1/4 supersymmetric when
the electric field $E$ is taken to the critical value $E=1$.
Upon this, we have an inverse relationship between magnetic
field and flux density,
\begin{equation}
\Pi_z= R\frac{1}{B},
\end{equation}
and the energy density decomposes into two pieces nicely
\begin{equation}
{\cal E}= R\left(\frac{1}{B}+B\right).
\end{equation}
In this last expression, the first term represents the energy of
fundamental strings and the second represents D0 energy. Tension
of D2 itself disappeared completely.

One aspect of this state particularly relevant
for this note is that one can reduce size of
the supertube without affecting the fundamental string charge. Let
us fix the total electric flux
\begin{equation}
n_{{\rm F1}}=2\pi \Pi_z =\frac{2\pi R}{B},
\end{equation}
and take the small radius limit, where the energy per unit length
becomes
\begin{equation}\label{bpsu}
2\pi {\cal E}=n_{{\rm F1}}+\frac{(2\pi R)^2}{n_{{\rm F1}}}
=n_{{\rm F1}}+n_{{\rm F1}}B^2.
\end{equation}
So as we decrease amount of magnetic flux on D2, the cylindrical
configuration collapses to infinitesimal tubular configuration
which entirely consists of the fundamental string charge.

It was recently suggested that in the system of unstable D$p$-brane
decay, fundamental string may be realized as a limiting
configuration of tubular D$(p-1)$-branes with electric flux on
it~\cite{Se03bc}. Question of fundamental strings remains one
of more puzzling aspect in the study of unstable D-brane decay
\cite{Yi99hd,Be00xf,Gi00hf,Se00kd,Kleban,Gi02tv}, and very recently
possible roles of topological defects in this question
began to be addressed \cite{Ha02sk,Ha03qx}. Approaching it
from low energy dynamics, several nontrivial interaction between
topological defects were isolated, and it was found that tachyonic
kink configurations tend to attract nearby and parallel electric
flux lines \cite{Rey,Kw03qn} forming a bound state of D$(p-1)$ brane
and fundamental strings. The idea is to curl up such a bound state
into a cylindrical configuration with only one noncompact direction
and let the compact part shrink away to zero size.

Specific form of such tubular solution was obtained from
lifting \cite{Se99md}
the Callan-Maldacena solution of D$(p-1)$ brane ~\cite{CM,Gi97xz}
to that of tachyon effective action on unstable D$p$-brane
and concentrating on the spike part of the configuration.
Such a combination involves vanishing radius of the tube, which
makes it fairly singular.
In this note, we consider the case of unstable D3 brane and
consider a family of solutions representing supertube. These may
be regarded as a thickened version of such tubular fundamental string
charge carrier and may be used as regulated and stabilized solution.

The D$(p-1)$ branes are also typically represented by infinitesimally thin 
domain wall solutions~\cite{Sen}.
Nevertheless, smooth kink solutions are known and, in the theory of 
tachyon, are represented by an infinite
array of kinks and anti-kinks to be interpreted as 
D-brane-anti-${{\rm D}}$-brane pairs~\cite{La03zr,Ki03in,Br03rs}.
When the tachyon couples to Born-Infeld gauge field,
spectra of regular extended objects become rich~\cite{Ki03in, 5092}.
Some of these smooth solutions are interpreted as thickened D-branes
or bound states of D-brane and F1. 
Topological kink solution with orthogonal and uniform critical electric flux
appears to enjoy BPS properties.
For infinite array of kinks-anti-kinks 
and a topological kink with orthogonal and uniform electric flux,
exact boundary states were also found~\cite{Se03bc}.
Naturally, this prompts 
us to ask whether there exists a regulated and thickened version of 
supertube-like solutions, which of course should reproduce genuine 
supertube configurations in the zero thickness limit.

\vskip 5mm

Section 2 will introduce the setup and solve for supertube-like
configuration with ``critical'' electric field. With a specific
choice of potential, we find an analytic and smooth solutions
of coaxial array of tubes with D0 and fundamental string charge.
Surprisingly, this family of solutions seemingly saturate a
BPS-like energy formula, and behave much like supertubes, except
the D2 worldvolume is thickened. 
Taking an infinitesimally 
thin limit of this solution, we find singular supertube solutions 
of arbitrary radius which match up with expected properties of
supertubes precisely. We close the section with brief comments
on relevance of these solutions in the
context of fundamental string formation.

Section 3 addresses the question of turning on more electric fields
and point out that boosting the tubular solution lengthwise will
tilt the flux such that fundamental string appears winding helically
along the tube. Section 4 considers a different class of ansatz
involving Dirac monopole in gauge sector. We show that the only
static solution of such kind involves strict vacuum $T=\infty$
so that the monopole-like configuration cannot lead to any
physical effect. The appendix explores supertube ansatz with
noncritical $E<1$ case. We close with summary.\footnote{As 
this work was finished, a related work discussing singular
supertube solution appeared on the net \cite{Mart}. }

\setcounter{equation}{0}
\section{Supertubes from Unstable D3-Brane}

We begin this section with introducing
effective tachyon action for unstable
D3-brane~\cite{Se99xm,Se99md,Ga00tr,Be00dq,Kl00iy,Ar00pe}
\begin{equation}\label{fa}
S= -{\cal T}_3 \int d^4x\; V(T) \sqrt{-\det (g_{\mu\nu} +
\partial_\mu T\partial_\nu T + F_{\mu\nu})}\, .
\end{equation}
This sort of behavior with potential multiplying the kinetic
term seems generic and in particular predicted by boundary string
field theory approach \cite{Ge00zp,Minahan}.
For most of this note, we will adopt a convenient form of $V(T)$
~\cite{BLW,Ki03he,Le03db,La03zr}, which has been derived
from open string theory~\cite{Ku03er} recently,
\begin{equation}\label{V3}
V(T)=\frac{1}{\cosh \left(T/T_{0}\right)},
\end{equation}
where $T_{0}$ is $\sqrt{2}$ for the non-BPS D-brane in the superstring
and 2 for the bosonic string, in the string unit.

We are interested in tubular configurations embedded in flat D3-brane,
and will work in the cylindrical coordinate system,
\begin{equation}
x^\mu = (t,x^{i})=(t,z,r,\theta),\qquad
ds^2 = -dt^2 + dz^2 + dr^2 + r^2 d\theta^2 .
\end{equation}
Both the tachyon field and the electromagnetic fields are assumed to
depend only on the radial coordinate $r$, and we employ ansatz for the
electromagnetic fields consistent with supertube solution
\begin{eqnarray}
&&T = T(r), \\
&& F_{0r}=F_{0\theta}=F_{r\theta}= F_{zr}=0.
\label{emf}
\end{eqnarray}
With the assumption of time independence, behavior of
$E_z$ and $F_{\theta z}$ is
determined entirely by the Bianchi identity, which gives
\begin{equation}
E_z=E, \qquad F_{\theta z}/r = B(r)=\alpha/r,
\end{equation}
with constant $E$ and $\alpha$.\footnote{This ansatz fails Bianchi identity
$\nabla\cdot B=0 $ at origin.}
Since we have
\begin{eqnarray}\label{X}
- X \equiv -\det (g_{\mu\nu} +\partial_\mu T\partial_\nu T + F_{\mu\nu})
= \left[(1-E^2)r^2+\alpha^{2}\right]( 1+ {T'}^2),
\end{eqnarray}
the low energy effective action simplifies drastically
\begin{equation}
-{\cal T}_3\int dtdzd\theta\;\int dr\;V(T)
\sqrt{(1-E^2)r^2+\alpha^2}\sqrt{1+T'^2}\, ,
\end{equation}
where the prime ${}'$ denotes differentiation with respect to the radial
coordinate $r$.

\subsection{Tachyon Tube}

Since we are primarily concerned with supertube-like solutions,
we may as well take $E=1$ for start and work with the action
\begin{equation}
-\alpha{\cal T}_3\int dtdzd\theta\;\int dr\;V(T)\sqrt{1+T'^2} \, .
\end{equation}
We will come back in later part of the note for more general  solution.
Then the effective action above maps to that of a simple mechanical
system with conserved ``energy'' by imagining $r$ as
``time'', which immediately  gives us the following integral of motion,
\begin{equation}\label{ineq}
\frac{V(T)}{\sqrt{1+T'^2}}=\beta\alpha/{\cal T}_3 ,
\end{equation}
where we introduced an arbitrary integration constant $\beta$.
Equivalently this integral of motion can be obtained from
the energy-momentum conservation of the full action.
Eq.~(\ref{ineq}) may be rewritten as
\begin{equation}\label{feq}
\frac{1}{2}T'^{2}+U(T)=-\frac12 ,
\end{equation}
with
\begin{equation}\label{upo}
U(T)=-\frac{1}{2(\alpha\beta)^{2}}\left[{\cal T}_{3}V(T)\right]^{2}.
\end{equation}
Note that, in the limit of vanishing magnetic field $\alpha=0$, the equation
becomes $V(T)=0$ and it allows only trivial vacuum solution $T(r)=\pm\infty$.

A nontrivial solution exists for $(\alpha\beta/{\cal T}_{3})^{2}<1$
and is a coaxial  array of tubular configuration where $T$ oscillates
as function of $r$. For instance, under the specific form of tachyon
potential (\ref{V3}), we have an exact solution
\begin{equation}\label{tub}
T(r)=-T_{0}\, {\rm arcsinh}\left[\sqrt{\left(
\frac{{\cal T}_{3}}{\alpha\beta}\right)^{2}-1}
\,\cos \left(\frac{r+r_0}{T_{0}}\right)\right].
\end{equation}
Note that this represents a coaxial array of tubular D2 and anti-D2
since the solution involves kink and anti-kink pair along every $\Delta r
=2\pi T$ \cite{horava}.
For the moment, we shall impose regularity
of $T$ field at origin by requiring $r_0=m\pi T_{0}$ where
$m$ is an integer.
Note that the periodicity of the solution, $2\pi T_{0}$, is
independent of the value of $(\alpha\beta/{\cal T}_{3})^{2}$.
This is a consequence of the particular form of the potential (\ref{V3})
we adopted.

\begin{figure}[t]
\centerline{\epsfig{figure=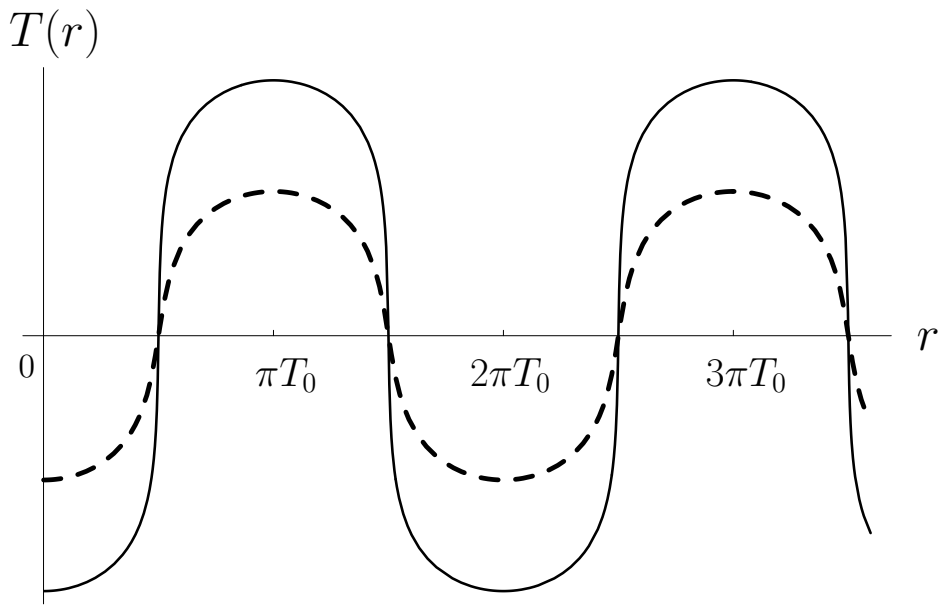,height=80mm}}
\caption{Plot of $T(r)$ with $r_0=0$, $T_0 =1$, $\alpha =1$, and $E = 1$:
(i) solid line for $\alpha\beta/{\cal T}_{3}=0.001$, and (ii)
dashed line for $\alpha\beta/{\cal T}_{3}=0.1$. } \label{fig1}
\end{figure}

For the obtained tubular objects, the following physical momenta
vanish,
\begin{equation}
T_{0z}=0, \qquad T_{0r}=0,
\end{equation}
and all off-diagonal components of stress-tensor 
vanish as would be expected of a stationary solution.
All nonvanishing conserved densities are dictated by a single quantity
\begin{equation}
\Sigma(r)\equiv \frac{{\cal T}_3}{\alpha}V(T)\sqrt{1+T'^2}=\beta (1+T'^2)
\end{equation}
which, for the particular form of the potential (\ref{V3}),
has an explicit expression
\begin{equation}
\Sigma(r)\rightarrow \beta
\left\{1+\frac{\sin^{2}(r/T_{0})}{\left[({\cal T}_{3}/\alpha\beta)^{2}
-1\right]^{-1}+\cos^{2}(r/T_{0})}\right\}.
\end{equation}
The energy and the electric
flux in infinitesimal $rdrd\theta$ are written as
\begin{eqnarray}
{\cal E} \,drd\theta&=& (r^2+\alpha^2)\Sigma(r)\,drd\theta ,\\
\Pi \, dr d\theta &=& r^2\Sigma(r)\,drd\theta ,
\end{eqnarray}
per unit length along $z$. For these quantities, we can see that the
period along $r$ is actually half of that of $T(r)$. This can be
understood by realizing that solution represents a coaxial
array of tubular D2 and anti-D2 branes with worldvolume gauge
fields turned on. While a single isolated D2 in vacuum is given
by singular kink, it is also known that modification of boundary condition
may thicken the kink into  a smooth thickened solution
\cite{Ki03in,Br03rs,5092}. Each period of $T$
contains a pair of tubular domain wall and anti-domain wall, while the above
conserved quantities are insensitive on orientation of the wall.
Distinction between D2 and anti-D2 can be seen from gradient of $T$ which
generates RR charge of D2 branes \cite{Se98sm,horava,Se99md}.

\begin{figure}[t]
\centerline{\epsfig{figure=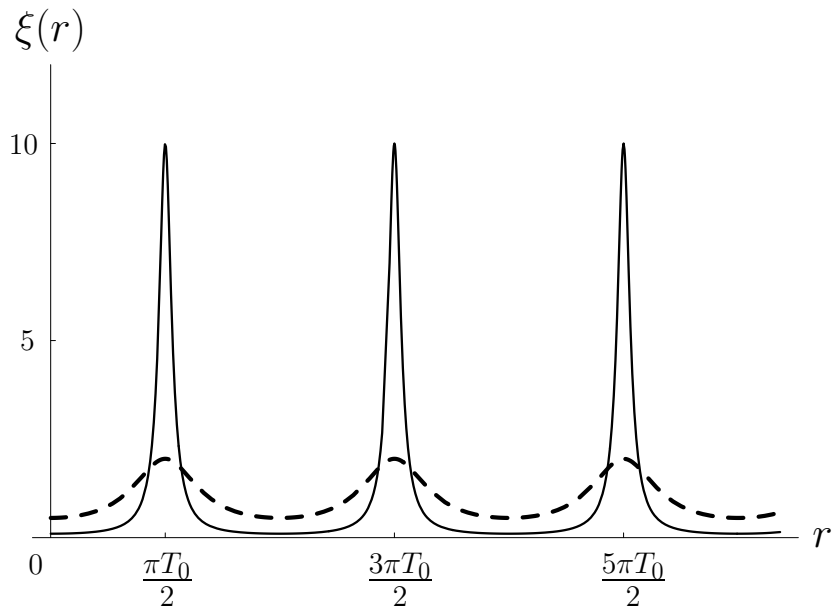,height=80mm}}
\caption{Plot of $\xi (r) \equiv {\cal E}(r) -\Pi(r)$ with
$r_0=0$, $T_0 =1$, $\alpha =1$, and $E_z = 1$:
(i) solid line for $\alpha\beta/{\cal T}_{3}=0.1$,
and dashed line for $\alpha\beta/{\cal T}_{3}=0.5$.}
\label{fig2}
\end{figure}

In order to identify the obtained tubular objects in
each half period $\Delta r=\pi T_0$, we compute the energy (tube tension)
per unit tube length
\begin{equation}\label{lene}
{\cal E}_{2}^{(n)}=2\pi\int_{(n-1)\pi T_{0}}^{n\pi T_{0}}dr\, {\cal E}
= 2\pi\beta \int_{(n-1)\pi T_{0}}^{n\pi T_0} \ dr \
  \frac{({\cal T}_3/\alpha\beta)^2 (r^2 + \alpha^2)}{
 1 +\left[({\cal T}_3/\alpha\beta)^2 -1 \right] \cos^2 (r/T_0)},
\end{equation}
and string charge per unit length
\begin{equation}\label{lsc}
Q_{{\rm F1}}^{(n)}=2\pi\int_{(n-1)\pi T_{0}}^{n\pi T_{0}}dr \,\Pi
=2\pi\beta \int_{(n-1)\pi T_{0}}^{n\pi T_0} \ dr \
  \frac{({\cal T}_3/\alpha\beta)^2 \ r^2 }{
 1 +\left[({\cal T}_3/\alpha\beta)^2 -1 \right] \cos^2 (r/T_0)}.
\end{equation}
In addition, each unit tube (or antitube) carries angular momentum
per unit length
\begin{equation}\label{ang}
L^{(n)}
=-2\pi\alpha\beta\int_{(n-1)\pi T_{0}}^{n\pi T_{0}}dr
\frac{({\cal T}_3/\alpha\beta)^2 \ r^2 }{
 1 +\left[({\cal T}_3/\alpha\beta)^2 -1 \right] \cos^2 (r/T_0)}.
\end{equation}
A salient point is that the difference ${\cal E}-\Pi$
is quite simple and can be identified as D0-brane charge.
In terms of formulae above, we define $Q_{{\rm D0}}^{(n)}$ as
\begin{equation}\label{ld0}
{\cal E}_{2}^{(n)}=
Q_{{\rm D0}}^{(n)}+Q_{{\rm F1}}^{(n)}.
\end{equation}
For the specific form of the tachyon potential (\ref{V3}), it
coincides exactly with the D0-brane charge before introducing
tachyon profile
\begin{equation}\label{qd}
Q_{{\rm D0}}^{(n)}=
2\pi\beta\alpha^2 \int_{(n-1)\pi T_{0}}^{n\pi T_0} \ dr \
  \frac{({\cal T}_3/\alpha\beta)^2 }{1 +\left[({\cal T}_3 /
  \alpha\beta)^2 -1 \right] \cos^2 (r/T_0)}
=2\pi^{2} \alpha T_0 {\cal T}_3 ,
\end{equation}
regardless of value of $\beta$. Thus, much like the case of supertube we
introduced earlier, the tension of unit tachyon tube (or antitube)
system decomposes linearly into two additive pieces.

Furthermore $Q_{\rm D0}$, as the notation suggests,  happens to be the
energy density expected of D0-branes. To see this, we need to note that
\begin{equation}
{\cal T}_2=\pi T_0{\cal T}_3 
\end{equation}
is the quantity that becomes  tension of the D2-brane
in this low energy theory. On the other hand,
$2\pi\alpha$ is the net magnetic flux on the D2 per unit $z$ length.
Product of the two generates D0 charge in spacetime \cite{Li}.

However, we must caution the reader against taking this fact too
seriously. Most likely, this miracle happened only because of
the particular form of the potential (\ref{V3}) adopted above,
which is justified only in certain limit \cite{Ku03er}.
A priori, there seems to be no rationale why the integral that defines
$Q_{\rm D0}^{(n)}$ (\ref{qd}) must be independent
of $\beta$, for more general form of potential. Dependence on $\beta$
will result in an anomalous dependence on the electric flux density and
ruin the BPS-like energy formula. In such general
case, one must take the limit of infinitesimal thickness of the
domain wall to find such a simple behavior, which is the subject
of next part. Nevertheless, as long as we are working with such a
particular potential, this gives an interesting option for handling
supertube-like configurations.

\subsection{Thin Supertube}

When we take a limit of $\beta\rightarrow 0$, the configuration gets
squeezed near $r=(n-1/2)\pi T_0$ for each positive integer $n$.
Every localized piece becomes sharply peaked at each site of the
tachyon tube (or antitube) so that
\begin{equation}
\lim_{\beta \to 0} \beta (1+{T'}^2)
= \frac{\pi {\cal T}_3 T_0}{\alpha}
\  \sum_{n=1}^{\infty}\delta \left(r-(n-1/2)\pi T_0\right),
\end{equation}
as shown in Fig.~\ref{fig2}.

This singular form of solution also implies that value of $T$
lies at $\pm \infty$ away from these centers, and neither the value of
$T'$ nor $B$ enters the dynamics directly except for discrete positions of
tube (antitube). In such limit, we have no reason whatsoever to insist upon
regularity at the origin. We thus have more general
{\it singular} solution such that
\begin{equation}
\lim_{\beta \to 0} \beta (1+{T'}^2)
= \frac{\pi {\cal T}_3 T_0}{\alpha}
\  \sum_{n=1}^{\infty}\delta \left(r+r_0-(n-1/2)\pi T_0\right).
\end{equation}
Without loss of generality we may consider values of $r_0$ such that
$\pi T_0/2>r_0\ge 0$. Correspondingly, we have
\begin{eqnarray}
{\cal E}_2^{(n)} &=&\frac{2\pi }{\alpha}
\left[\left((n - 1/2)\pi T_0 -r_0\right)^2 + \alpha^2\right]
\times\pi{\cal T}_3 T_0,
\label{t2n}\\
Q_{{\rm F1}}^{(n)} &=& \frac{2\pi}
{\alpha}\left((n - 1/2)\pi T_0-r_0\right)^2\times\pi{\cal T}_3 T_0,\label{qn}\\
Q_{{\rm D0}}^{(n)} &=& {2\pi\alpha}\times\pi{\cal T}_3 T_0 ,
\end{eqnarray}
and the accompanying angular momenta,
\begin{equation}
L^{(n)} = -2\pi
\left( (n-1/2)\pi T_{0} - r_0\right)^2\times\pi{\cal T}_3 T_0 .
\end{equation}
Though distribution of the D0 charge and its magnetic field profile
are singular, induced energy contribution is finite as shown in Eq.~(\ref{t2n}).
Note that Eq.~(\ref{ld0}) and Eq.~(\ref{qn}) satisfy the relation between
radial size of $n$-th tube $R_{n}=(n-1/2)\pi T_{0}-r_0$ and charges such as
$R_{n}=\sqrt{Q_{{\rm F1}}^{(n)}Q_{{\rm D0}}^{(n)}}/2\pi^{2}T_{0}{\cal T}_{3}$.

Different choice of the tachyon potential will modify the universal
function $\Sigma(r)$, but the limiting form will again involve
delta-functions. For the following discussion we will consider
more general form of the potential, and retains the requirement
that the potential decays exponentially $\sim e^{-|T|}$ asymptotically.
Other requirements are that $V(T)$ is monotonically decreasing
in $|T|$ and that $V(0)=1$.

For convenience, take a half period of $T(r)$ that solves
\begin{equation}
T'(r)=\sqrt{({\cal T}_3 V(T)/\beta\alpha)^2-1}\, .
\end{equation}
$T'(r)$ then starts out with value zero and begins to increase slowly
as long as $V(T)$ value remains exponentially small in $T$. Conversely
when $T'$ remains small $V(T)$ value must remain small also.
Let us take any of one of half period such that $T(r)$ starts at large
negative value $-T_{{\rm max}}$, where $V(\pm T_{{\rm max}})=\beta\alpha/{\cal T}_3$
for the rest of discussion.
So the half period is determined by the integral
\begin{equation}
\Delta r_{\rm half}=\int_{-T_{{\rm max}}}^{T_{{\rm max}}}
\frac{dT}{\sqrt{({\cal T}_3 V(T)/\beta\alpha)^2-1}}\, .
\end{equation}
At some points near the middle of this half period, $T$ begins to approach
$-{\sf T}_0$ where ${\sf T}_0$ now refers to characteristic scale where 
$V(T)$ is of order
1. In this region $T'(r)$ is very large, achieving a maximum value at
$r=R$ such that $T(R)=0$. This means that the interval in $r$ where
$T$ is comparable to ${\sf T}_0$ is very short.

Recall that the function $\Sigma$ is given by
\begin{equation}
\Sigma(r)=\frac{{\cal T}_3}{\alpha}V(T)\sqrt{1+T'^2} \, .
\end{equation}
We wish to approximate this expression by
\begin{equation}
\Sigma(r)\sim\Sigma_0=\frac{{\cal T}_3}{\alpha}V(T)T' ,
\end{equation}
for monotonic and increasing $T(r)$. Of course this is possible only
if $T'$ is very large. In the region where $T'$ is of order one,
the equation for $T$ above tells us that we also must have,
\begin{equation}
{\cal T}_3 V(T)/\beta\alpha \sim 1 ,
\end{equation}
meaning that $V(T)$ is still of order $\beta$. Since we are
taking the limit of $\beta\rightarrow 0$, this allows us to
use the approximate formula $\Sigma_0$ in place of $\Sigma$.
Within this half period, assuming that the latter is finite,
$\Sigma(r)$ reduces to
\begin{equation}
\lim_{\beta\rightarrow 0}\Sigma(r)\quad\rightarrow \quad
\frac{1}{\alpha}\times \left({\cal T}_3\int V(T)\,dT\right)\times
\delta(r-R).
\end{equation}
We have used the fact that the support of $\Sigma$, more precisely
the region of $r$ where $\Sigma$ remains larger than, say,
$\sim \beta^{1/3}$ vanishes rapidly as $\sim \beta^{1/3}$ when
we take $\beta \rightarrow 0$.

The quantity in the parenthesis
is  to be identified with ordinary BPS D2-brane tension
${\cal T}_2$ \cite{Sen}.
With this, D0 charge and F1 charge are respectively,
\begin{eqnarray}
Q_{\rm D0}&=&2\pi\alpha {\cal T}_2 ,\\
Q_{\rm F1}&=&\frac{2\pi R^2}{\alpha} {\cal T}_2 ,
\end{eqnarray}
and the energy is the sum of the two
\begin{equation}\label{sum}
{\cal E}_2= Q_{\rm D0}+Q_{\rm F1}.
\end{equation}
In order to compare this to ordinary supertube, we need to restore
the factor of D2-brane tension, ${\cal T}_2$, in energy formulae of
section 1 and identify $\alpha$ with $BR$.

Therefore, the supertube can be understood as that in singular
zero-thickness limit of the obtained unit tachyon tube (or antitube) 
solution.
The whole configuration we obtained is interpreted as coaxial array of
tube-antitube. Once we take $\beta=0$ limit, on the other hand,
each and every tube becomes a complete solution on its own.

\subsection{Tachyon Tube and Tubular Fundamental String}

As we saw earlier in this note, supertube provides an interesting method
of blowing up a long fundamental string. While this is not much of advantage
in ordinary supersymmetric string theory context, it may prove more
useful when we consider how fundamental string forms from decay of
unstable D-branes.

One of more intriguing scenarios that emerged recently involve string flux
sitting at top of tachyon potential. However, no such stable solution exists
and one must take a double scaling limit of cylindrical domain wall to
reach such a state.  That is, take an infinitely thin domain wall with
fluxes on it and  then take infinitesimal radius limit of a
cylinder thereof. The string flux is imagined to be embedded in
this infinitesimal cylinder. Because of such a double scaling limit,
the object is a bit difficult to handle.

A priori,  it is not obvious whether,  in the absence of D$(p-1)$
branes nearby, there is really any tangible difference
between such configurations and a tight collection of string fluid \cite{Gi00hf}
sitting at some homogeneous vacuum.
As was point out in Ref.~\cite{Se03bc}, the dynamics of the two
will follow exactly the same, Nambu-Goto-like,
equation of motion \cite{Gi00hf,Se00kd,Nielsen} away from D$(p-1)$ brane.
One motivation for the current work is to provide a setting where
more precise question about such stringy configurations can be asked.

By adding magnetic field to the tubular configuration, we essentially
obviated one of the two scalings, and found a
stable string-like configuration of small but finite tubular radius.
Magnetic field strength necessary scales with the radius of the tube,
so one can take smaller and smaller size tube by taking small magnetic
field limit.
For specific form of the potential, furthermore, the unit tachyon tube
before taking thin supertube limit provides additional control, given
that it also
satisfies a BPS-type relation (\ref{ld0}) and is given by exact
and smooth solution (\ref{tub}). Existence of such a static and smooth
profile in closed form may come in handy, at least for the purpose of
having a set of solutions that approaches infinitesimally small tubular
solution.\footnote{This is not to say that these are a degenerate
one-parameter family of solutions associated with a supertube.
For smooth tube solution, the net electric flux in each unit tube
depends on the parameter $\beta$ while the D0 charge remains unchanged
for the particular form of the potential. What remains unchanged
is that the energy gets contribution from that of fundamental strings
and D0's only.}

While we concentrated on cylinders with
circular cross section, it is well known that a supertube  can be
deformed to have an arbitrary cross section \cite{tube,Mateos,karch}. 
In the present context,
this results in huge degeneracy which gives at best arbitrarily shaped
distribution of fundamental string charges, much as in the case of
string fluid in real vacuum. As was also pointed out previously, therefore,
this alternate picture of fundamental string is so far unable to resolve
the usual degeneracy problem \cite{Be00xf,Gi00hf,Kw03qn}, or equivalently
the lack of confinement mechanism, at the level of classical effective theory.
It remains to be seen how and at what level this degeneracy is
lifted \cite{Yi99hd,Be00xf,Kleban,Se03bc}.

\setcounter{equation}{0}
\section{Boosted Tachyon Tube with Helical String Flux}

One might ask whether more general ansatz leads to new kind of
tubular solutions. One of more obvious variation we can try
is to turn on more electromagnetic field. In fact there is
a simple generalization of the ansatz that brings us back
to an equally simple integrable equation for $T$. Consider
turning on $E_\theta$ in addition to $E_z$
\begin{equation}
F=E_z dt\wedge dz+E_\theta dt \wedge d\theta+ \tilde\alpha
d\theta\wedge dz .
\end{equation}
With static field strength of $r$ dependence alone,
Bianchi identity demands that both $E_\theta $ and
$E_z$ are constants.

Note that  $E_\theta \neq 0$ will generate
fundamental string charge along angular direction in addition to
the one along $z$ direction, so the electric flux will have
a helical shape along the tube. More precisely, the electric flux
along $z$ and $\theta$ directions has a relation
\begin{equation}
{\Pi_\theta} = \frac{E_\theta}{E_z} \Pi_z,
\end{equation}
where we define $\Pi^\theta \equiv \delta S/\delta E_\theta = \Pi_\theta/r^2$,
and $\Pi^z \equiv \delta S/\delta E_z$ in $A_0=0$ gauge.
The effective action for the tachyon field $T=T(r)$
becomes
\begin{equation}
-{\cal T}_3\int dtdzd\theta\;
\int dr\;V(T)\sqrt{(1-E_z^2)r^2+\left(\tilde\alpha^2
-E_\theta^2\right)}\sqrt{1+T'^2}\, .
\end{equation}
As before, this becomes integrable once we impose the condition of
critical electric field, $E_z^2=1$, since
the Lagrangian will then have no explicit $r$-dependence. Furthermore
it is easy to show that, within cylindrical symmetry ansatz, this is 
the only case where simple first order integral of motion is
possible.

In fact,
the form of the action is identical to the case with $E_\theta=0$
above if we identify
\begin{equation}\label{alp}
\sqrt{\alpha^2} = \sqrt{\tilde \alpha^2-E_\theta^2}\, .
\end{equation}
Note in particular that sensible solution will appear only if
$\tilde\alpha^2> E_\theta^2$.
One can understand this by realizing that the two cases are in
fact related by a Lorentz boost along $z$ direction. Starting with
the simplest case,
\begin{equation}
F=E_z dt\wedge dz+ \alpha d\theta\wedge  dz ,
\end{equation}
one may boost this field strength along $z$ with velocity $v$,
\begin{equation}
F=E_z dt\wedge dz+  \frac{\alpha v}{\sqrt{1-v^2}}
\,dt\wedge d\theta+\frac{\alpha }{\sqrt{1-v^2}} \,d\theta\wedge dz .
\end{equation}
We then may identify
\begin{eqnarray}
\tilde \alpha=\frac{\alpha }{\sqrt{1-v^2}},\qquad
E_\theta =\frac{\alpha v}{\sqrt{1-v^2}} ,
\end{eqnarray}
which also gives Eq.~(\ref{alp})
completing the assertion that turning on $E_\theta$ is equivalent to
boosting and appropriate rescaling of $B$. Of course $T$ is a
scalar field and so $T'$ is also invariant under the Lorentz boost
along $z$ direction.

Emergence of winding electric flux on D2, or equivalently winding
fundamental string charge, upon a boost may be surprising. After
all, the same will happen even if the D2-brane were wound on
a topological circle, which seems to say that boosting generated net
conserved string charge along the topological circle direction.
This unexpected effect may be understood once we realize that
worldvolume magnetic field also couples NS-NS field $B_{\mu\nu}$.
Although it does not generate net string charge, the
worldvolume magnetic field generates a spacelike current
associated with string charge.\footnote{We thank
Kimyeong Lee on this point.} All that happen is that the Lorentz
boost tilts this spacelike current toward time direction a bit,
which effectively generates density of string charge.

\setcounter{equation}{0}
\section{Is There a Magnetic Monopole in an Unstable D3?}

With a minimal assumption on nontrivial field content, the effective
action was shown above to be of the form,
\begin{equation}
\int dr\; V(T)R(r)\sqrt{1+T'^2}
\end{equation}
with
\begin{equation}\label{stb}
R(r)=\sqrt{(1-E^2)r^2+\alpha^2}
\end{equation}
for the supertube case with constant $rB=\alpha$. Reality of
the Born-Infeld action requires, $E^2\le 1$, since otherwise the
action will become purely imaginary at asymptotic region.

This sort of
effective action can be used to explore another kind of
configuration, namely a spherically symmetry magnetic monopole.
The latter should have
\begin{equation}
F=Q\sin\theta d\theta\wedge d\phi
\end{equation}
which gives an effective action of the above form with
\begin{equation}\label{mpl}
R(r)=\sqrt{r^4+Q^2}.
\end{equation}
Of course $r$ in Eq.~(\ref{mpl}) is the radial coordinate of spherical
coordinate system.

Unlike the case of Maxwell theory, the singularity
of this Dirac monopole does not cost infinite amount of energy, and
one might wonder if there exist a Dirac monopole-like solution of finite
mass. Existence of such a solution is unlikely, given that no obvious
stringy interpretation is available. Here we will confirm this
expectation by showing that the above class of
action does not possess a regular and
isolated solution that approaches vacuum at large radius.

If there is an isolated solution with such charges, $T(r)$ must approach
$\pm \infty$ at large $r$. Since the system is invariant under $T
\rightarrow -T$, take positive infinity for the asymptotic value of
$T$. Far away from origin, then, the potential $V$ approaches $e^{-T}$
exponentially fast, and we could ask whether there is a proper
solution with $V=e^{-T}$ and $T(\infty)=+\infty$. Euler-Lagrange
equation of motion reduces to
\begin{equation}\label{ele}
\frac{d}{dr}\left(\frac{RT'}{\sqrt{1+T'^2}}\right)
+\frac{R}{\sqrt{1+T'^2}} =0,
\end{equation}
where we used $\partial_T e^{-T}=-e^{-T}$. Defining $f$ as
\begin{equation}
f(r)\equiv \frac{T'}{\sqrt{1+T'^2}},
\end{equation}
the equation of motion (\ref{ele}) is now
\begin{equation}
\frac{d}{dr}\left(Rf\right)
+R\sqrt{1-f^2} =0.
\end{equation}
Because of the relative positive sign, this first order equation
always drives $f$ toward $-1$ from above as $r\rightarrow \infty$
for both choices of $R(r)$. Near $r=\infty$, we find
\begin{equation}
f=-1+\frac{c}{2r^2}+\cdots,
\end{equation}
where $c=1$ for supertube ansatz (\ref{stb}) and $c =4$ for monopole
ansatz (\ref{mpl}). The corresponding $T$ is, on the other hand,
\begin{equation}
T(r) = \int dr  \;\frac{f}{\sqrt{1-f^2}}=
-\int dr\;\frac{r}{\sqrt{c}}+\cdots
=-\frac{r^2}{2\sqrt{c}}+\cdots
\end{equation}
which is inconsistent with original assumption $T\rightarrow +\infty$.
Thus there exists no isolated supertube or monopole solution such that
$T$ at finite $r$ is finite.

Thus only acceptable solutions that approach vacuum at infinity
are singular. In case of supertube ansatz, it corresponds to any one
of the coaxial tubes in $\beta \rightarrow 0$ limit in section 2.
In case of monopole,
it corresponds to $T=\infty$ everywhere with a Dirac monopole in the
gauge sector. In the latter case, the Dirac monopole does not generate any
measurable quantity, since all conserved quantum numbers we considered
above vanish for this configuration. Only if there is an electric charge
that couples to the gauge field directly (say, via minimal vector-current
coupling), can there be physical consequence of this monopole solution.
In the context of unstable D-brane, no such charged particle is
known to exist.

\section{Summary}

In this note we considered static tubular solution in unstable D3-brane
decay with an ansatz that involves both D0 charge and fundamental string
charge. A series of both smooth and singular solutions are obtained
and compared to supertube configuration of D2. While for a specific choice
of potential there exists a smooth BPS-like solution, for generic choice
of potential, one must take a singular limit to find supertube solution.
After a brief comment on possible relevant of such solutions in study
of fundamental string formation, we turned to other exotic solutions.
Finally we show that magnetic monopole solution does not
exist as a sensible finite energy solution.

\section*{Acknowledgements}
We would like to thank K. Hashimoto,  and A. Sen for helpful
discussions. Y.K. was indebted to the late Youngjai Kiem who introduced and
taught the topic of supertube in spring, 2001.
C.K. is supported by Korea Research Foundation Grant
KRF-2002-070-C00025. This work is the result of
research activities (Astrophysical Research
Center for the Structure and Evolution of the Cosmos (ARCSEC))
supported by Korea Science $\&$ Engineering Foundation(Y.K. and O.K.).
P.Y. is supported
in part by Korea Research Foundation (KRF) Grant KRF-2002-070-C00022.

\setcounter{equation}{0}
\renewcommand{\theequation}{A.\arabic{equation}}
\section*{Appendix: Tachyon Tubes with Noncritical Electric Field}

In the previous sections we dealt with tubular objects with critical
value of the electric field $E^{2}=1$. In this appendix let us consider
arbitrary value of the $z$-component of the electric field $E^{2}< 1$
and investigate effect of it. As we saw above, isolated solution does
not exist. Rather we shall find that the coaxial structure found
in $E^2=1$ case still persist, but with some quantitative modification.
Below we will study this case.

Finding solution is a matter of finding
classical solution of the following reduced Lagrangian
\begin{equation}
-V(T)\sqrt{(1-E^2)r^2+\alpha^2}\sqrt{1+T'^2}\, ,
\end{equation}
and we have the following second order equation to solve.
\begin{equation}\label{TE2}
\frac{ T''}{ 1 + {T'}^2}
+\frac{(1-E^2)r T'}{(1-E^2) r^2 + \alpha^2}
=\frac{1}{V}\frac{dV}{dT}.
\end{equation}Alternatively,
we can get the same equation from $r$-component of energy-momentum
conservation law, using
\begin{eqnarray}
T_{rr} &=& - {\cal T}_3\frac{V}{\sqrt{-X} \ r}\left[ r^2(1- E^2)
+ \alpha^2\right], \\
T_{\theta\theta}&=& - {\cal T}_3\frac{Vr^3}{\sqrt{-X}}(1+ {T'}^2) ( 1-
E^2).
\label{tth}
\end{eqnarray}
Since we kept the orthogonality between electric and magnetic fields,
physical electric flux is such that
\begin{eqnarray}\label{Pi}
\Pi_{r}=\Pi_{\theta}=0,~~\Pi(\equiv\Pi_{z}) ={\cal
T}_{3}\frac{Vr^2 E }{\sqrt{-X}}(1 + {T'}^2),
\end{eqnarray}
where $-X$ is given in Eq.~(\ref{X}).
$\Pi$ in Eq.~(\ref{Pi}) contains a factor of $\sqrt{-X}$, so we can see
that $E^2$ larger than 1 is not acceptable
physically, if we wish to maintain reality of the action.
Thus we will consider $E^2<1$ from now on.

\begin{figure}
\centerline{\epsfig{figure=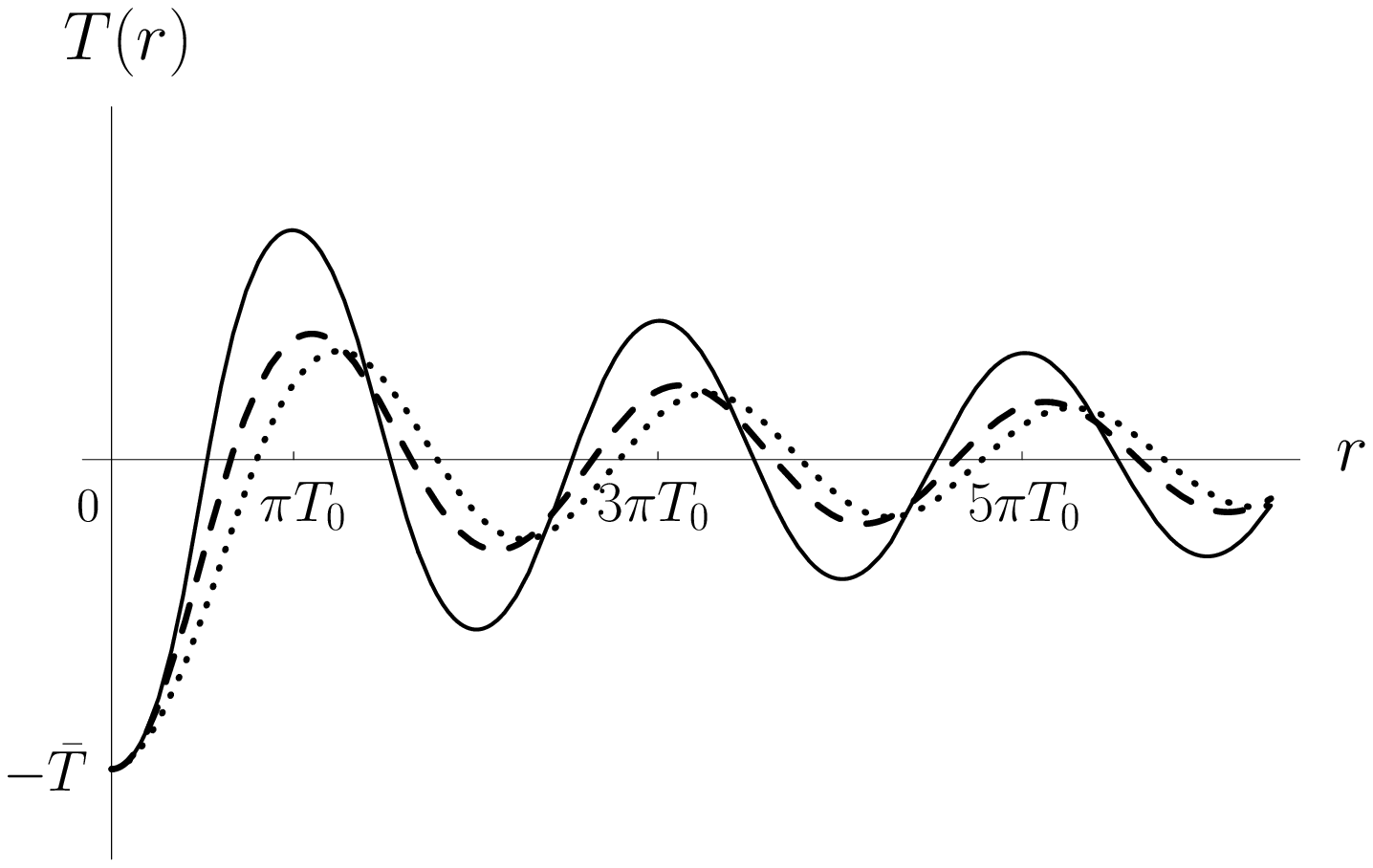,height=8cm}}
\caption{Plots of $T(r)$ with common initial value $-\bar{T}$:
(i) solid line for $\alpha^{2}/(1-E^{2})=10$, (ii) dashed line for
$\alpha^2/(1-E^{2})=1$, and (iii) dotted line for
$\alpha^2/(1-E^{2})= 0.1$.} \label{fig3}
\end{figure}

With this, the second term of Eq.~(\ref{TE2})
gives always damping effect.
A natural boundary condition at the origin is again to assume $T'(0)=0$.
Then the damping term of Eq.~(\ref{TE2}) does not affect choice of boundary
value of the tachyon field, so it can have any value similar to the free
parameter $\beta$ in Eq.~(\ref{ineq}) for the case of $E^{2}=1$, say
$T(0)=-{\bar T}$. Since the tachyon profile should be
nonsingular, power series expansion near the origin gives
\begin{equation}\label{tr0}
T(r)\approx -{\bar T} + a_{1} r^2 + a_{1}a_{2} r^4 + {\cal O}(r^6)
\end{equation}
with
\begin{eqnarray}
a_{1} &=&\frac{\tanh ({\bar T}/T_{0})}{2T_{0}},
\label{a1}\\
a_{2} &=& \left[ \frac{\tanh ({\bar T} /T_0)}{2T_0}\right]^{2}
 - \frac{1}{12 T_0^2} -\frac{1-E^2}{6\alpha^2}.
\label{a2}
\end{eqnarray}
Since the coefficient of $r^{2}$ term is determined only by the tachyon
potential and double derivative term of Eq.~(\ref{TE2}), it is independent
of the electric and magnetic fields. Its signature is opposite to that
of initial value $-{\bar T}$, which is consistent with monotonically
increasing (decreasing) behavior of tachyon tube (antitube).
In the limit of infinite ${\bar T}$, both second and third
terms in Eq.~(\ref{tr0})
remain finite and this implies possibility of zero thickness limit
of the tachyon tube with keeping the period finite.
As expected, electromagnetic contribution appears as decreased slope
in the $r^{4}$ term (\ref{a2}) and it implies increment of the period
(see Fig.~\ref{fig3}). Even in the limit of vanishing
$\alpha^{2}/(1-E^{2})$ but not exactly zero, the coefficient of the
second term of Eq.~(\ref{TE2}) has $1/r$ which decreases as $r$ increases.
For sufficiently large $r$, the approximated equation does not afford
the decaying solution to zero as $1/r^{p}$ form, so the critical or
overdamping solution seems unlikely.

Near $r_{c}$ with $T(r_{c})=0$, power series solution of Eq.~(\ref{TE2}) is
\begin{equation}\label{trc}
T(r)\approx b_{0}(r-r_{c})\left[1+b_{1}(r-r_{c})+b_{2}(r-r_{c})^{2}
+\cdots\right]
\end{equation}
with
\begin{eqnarray}
b_{1} &=&-\frac{r_{c}(1+b_{0}^{2})}{2\left(r_{c}^{2}+
\frac{\alpha^{2}}{1-E^{2}} \right)},
\label{b1}\\
b_{2} &=& \frac{1+b_{0}^{2}}{6}\left[
-\frac{1}{T_{0}^{2}}- \frac{1}{r_c^2 + \frac{\alpha^2}{1- E^2}}
+ \frac{3 r_c^2(1 +  b_0^2)}{\left(r_{c}^{2}
+\frac{\alpha^{2}}{1-E^{2}}\right)^{2}}\right],
\label{b2}
\end{eqnarray}
where $b_{0}$ and $r_{c}$ are determined by proper behavior near the origin.
Existence of even order terms in Eq.~(\ref{trc}), e.g., nonvanishing $b_{1}$
(\ref{b1}), breaks reflection symmetry about $r_{c}$ and stands for signal
of the damping due to the noncritical electric field.
In order to confirm the limit of step function for
sufficiently large $\alpha^{2}/(1-E^{2})$, let us compare the results
(\ref{trc}) with the critical case (\ref{tub}). Expansion of $T(r)$
in Eq.~(\ref{tub}) near $r_{c0}=(2n\pm\frac{1}{2})\pi T_{0}$:
\begin{equation}\label{trc0}
T(r)\approx\pm \sqrt{\left(\frac{{\cal T}_{3}}{\alpha\beta}\right)^{2}-1}
\, (r-r_{c0})\left[1- \frac{{\cal T}_3^2}{6 T_0^2 \alpha^2\beta^2}
(r- r_{c0})^2\right] + {\cal O}((r-r_{c0})^{5}).
\end{equation}
Taking the limit of critical electric field $E^{2}\rightarrow 1$
in Eq.~(\ref{trc}) with assumption $r_{c}\rightarrow r_{c0}$,
we observe as expected that
\begin{equation}\label{thi}
b_{0}\stackrel{E^{2}\rightarrow 1}{\longrightarrow}\pm
\sqrt{\left(\frac{{\cal T}_{3}}{\alpha\beta}\right)^{2}-1},
\qquad b_{0}b_{1}\stackrel{E^{2}\rightarrow 1}{\longrightarrow}0.
\end{equation}
The limiting value of $b_{0}b_{2}$ consistently reproduces
the cubic term in Eq.~(\ref{trc0}).
Therefore, near the critical $E$,
we can safely take zero thickness limit in a consistent way
with $\beta\rightarrow 0$ limit. On the other hand, once we take
$b_{0}\rightarrow \infty$ limit first with keeping $\alpha^{2}/(1-E^{2})$
finite, then $b_{0}b_{1}\sim -r_{c}b_{0}^{3}/\left[2(r_{c}^{2}
+\frac{\alpha^{2}}{1-E^{2}})\right]$ approaches $\mp\infty$ as far as
$r_{c}$ remains to be finite. Survival of even order terms implies asymmetry
of $T$ in the vicinity of $r_{c}$ and then derivative term $T'^{2}$ is not
expressed purely by $\delta$-function.

\end{document}